\documentclass[reprint,
superscriptaddress,
 amsmath,amssymb,
 aps,
]{revtex4-2}
\usepackage[utf8]{inputenc}
\usepackage{graphicx}
\usepackage{amsmath,amsthm,amssymb}
\usepackage{bm}
\usepackage{float}
\usepackage{lipsum}
\usepackage{comment}
\usepackage{tensor}
\usepackage{tikz-cd}
\usepackage{soul}

\usepackage{hyperref}
\hypersetup{
  colorlinks=true,        %
  linkcolor=blue,         %
  citecolor=cyan,         %
}

\usepackage{graphicx}
\usepackage{dcolumn}
\usepackage{bm}
\usepackage{color}

\begin{document}
\title{Stellar Objects from Quantum Gravity}
\author{Salvatore Samuele Sirletti}
\email{salvatoresamuele.sirletti@unife.it}
\affiliation{Dipartimento di Fisica e Scienze della Terra, Università degli Studi di Ferrara, via Saragat 1, I-44122 Ferrara, Italy}
\affiliation{INFN, Sezione di Ferrara, via Saragat 1, I-44122 Ferrara, Italy}
\affiliation{Kavli Institute for the Physics and Mathematics of the Universe (WPI), UTIAS, The University of Tokyo, Chiba 277-8583, Japan}
\affiliation{Dipartimento di Fisica, Università di Trento, Via Sommarive 14, 38123 Povo, Trento, Italy}
\author{Piero Nicolini}
\email{piero.nicolini@units.it}
\affiliation{Dipartimento di Fisica, Università degli Studi di Trieste, Strada Costiera, 11, 34151 Trieste TS, Italy}
\affiliation{Istituto Nazionale di Fisica
Nucleare (INFN), Sezione di Trieste,  Via Alfonso Valerio, 2, 34127 Trieste TS, Italy}
\affiliation{Institut f\"{u}r Theoretische Physik, Goethe-Universit\"{a}t, Frankfurt am Main, Germany}
\affiliation{Frankfurt Institute for Advanced Studies (FIAS), Frankfurt am Main, Germany}

\author{Mariafelicia De Laurentis}
\email{mariafelicia.delaurentis@unina.it}
\affiliation{Dipartimento di Fisica, Università degli Studi
di Napoli {}``Federico II'', Compl. Univ. di
Monte S. Angelo, Edificio G, Via Cinthia, I-80126, Napoli, Italy}
\affiliation{INFN Sezione  di Napoli, Compl. Univ. di
Monte S. Angelo, Edificio G, Via Cinthia, I-80126, Napoli, Italy.}

\date{\today}

\begin{abstract}

This paper explores the theoretical implications of quantum gravity by analyzing compact stellar objects, presenting three distinct models that serve as alternatives to traditional black holes. These models are characterized by their extreme compactness and incorporation of a quantum core, successfully avoiding the curvature singularities typically associated with classical general relativity. Central to these models is the noncommutative parameter, which plays a crucial role in determining stellar properties and enables the exploration of various astrophysical regimes. While pure Planckian effects pose significant challenges for observational detection, our findings suggest that lower energy scales may reveal exotic stellar objects with Earth and Sun-like masses that lack classical counterparts, potentially providing the first experimental evidence of non-classical gravity. We propose that the metrics derived in this study can be tested against known neutron stars, offering promising avenues for future research aimed at understanding the interplay between quantum effects in gravity and stellar evolution, ultimately enhancing our comprehension of the universe's fundamental properties.

\end{abstract}

\maketitle
\section{Introduction}
\label{intro}
 
Physics has made tremendous progress in recent years, both theoretically and experimentally, but it is still plagued by long-standing problems at its most fundamental level. For example, we have only a partial knowledge of the universe, its birth, and the early stages of its evolution \cite{Haw88}. 
To properly address the above questions, theoretical physicists have spent decades trying to formulate a quantum theory of gravity \cite{Nicolai13}.

Unfortunately, quantum gravity signals have not yet been detected because they appear near the Planck scale, i.e. $m_\mathrm{P}\sim 10^{28}$ eV, which is about 15 orders of magnitude higher than the most energetic event ever observed in a particle detector.
While unattainable in terms of energy, the Planck scale has another significance in terms of length: it corresponds to the size $l_\mathrm{P}\equiv 1/m_\mathrm{P}$ of the lightest black hole. This raises the general question of whether astrophysical observations can reveal quantum gravity better than particle physics experiments.

At this point, it should also be recalled that, in contrast to classical scenarios, there is no complete gravitational collapse in quantum gravity. At extreme energies $\sim m_\mathrm{P}$ and tiny length scales $\sim l_\mathrm{P}$, the repulsive quantum pressure due to virtual gravitaton exchange becomes dominant, stopping the collapse and forming a extreme Planckian density matter source \cite{Jun24}. 
Following this line of reasoning, since the 1960s there have been attempts to construct regular black holes by matching a local de Sitter core at the center with a Schwarzschild geometry around it \cite{Gli66,Sak66,Bar68,Dym92,AyG99c,MbK05,BrF06,Hay06}.
More recently, many authors have considered black holes as a viable testing ground for quantum gravity, but the predicted effects are too difficult to detect at current observational resolution \cite{DvG13b,Gid17,CMN24}.

With this in mind, we aim to explore the possibility of detectable quantum gravity effects by focusing on compact stellar objects rather than black holes. The scenario of a regular black hole is only one potential outcome of the final stage of gravitational collapse within the realm of quantum gravity. For specific values of its mass $M$, it has been conjectured that a star can collapse into a horizonless, non-singular compact object, which, since it is not predicted by classical general relativity, could serve as a smoking gun for quantum gravity.

Regarding our methodology, we note that noncommutative geometry (NCG)—which postulates that spacetime coordinates do not commute at very short distances—can be viewed as an effective theory for string theory and quantum gravity \cite{Seiberg:1999vs}. This framework has been successfully employed to construct regular black hole solutions \cite{Nicolini:2005vd,Nicolini:2008aj,Sir22} and can be extended to study other compact objects \cite{Sir22}, which is the main intent of this work. The advantage of this approach is its potential to yield model-independent conclusions, as it captures stringy effects similarly to other non-local deformations of gravity, based on T-duality \cite{NICOLINI2019134888} or generalized uncertainty principle arguments \cite{Carr:2015nqa}.

The paper is organized as follows: in section \ref{sec:RGOs}, after reviewing the basic properties of static stars with a regular, quantum core, we present three models, by varying some characteristics; in section \ref{sec:models}, we offer an analysis of typical length scales for the models of the previous section; in the concluding section \ref{sec:conclusions}, we comment on the possibility of future observations.

\section{Regular gravitating objects}
\label{sec:RGOs}

Since the initial work of Chadrasekhar \cite{Cha39}, stars have been a common research topic of astronomy and fundamental physics in trying to understand the properties of matter at high densities \cite{RuB69,ADLS85,Ris04}. 
However, the interplay between quantum gravity and compact objects is quite recent, probably dating back to 2001 when both the fuzzball \cite{Mat05} and the gravastar \cite{MaM23} proposals were presented, followed by the Planck star conjecture \cite{Rovelli:2014cta}.

Along the same lines, in the present paper we take advantage of spacetime metrics with short scale corrections due to noncomutative geometry. However, there are very important differences of our proposal with respect to the above mentioned models. We can group them as follows:
\begin{enumerate}
\item noncommutative geometry deforms the field propagators and allows the calculation of the virtual particle exchange static potential \cite{MMN11}: the regularity of spacetime is therefore a result of quantum gravity rather than a solution obtained with mathematical artifacts imposed by hand;
\item \label{item:ii} noncommutative geometry metrics describe either black holes or horizonless objects depending on the value of the mass: thus, the resulting stellar objects are not in opposition to black holes, but can coexist with them.
\end{enumerate}
Another important note concerns the symmetry of the metrics. Even if rotation is important, any quantum hair for black holes or any long-range non-classical effects for stars should already be visible in static solutions at the level of the modification of the Newtonian potential, $\phi(r)$. 

For the above reasons, in this initial work, we consider only static, spherically symmetric spacetimes whose line element is
   \begin{equation}
    ds^2 = -e^{2\Phi(r)} \left( 1-\frac{2m(r)}{r}\right) dt^2 +\frac{dr^2}{1-\frac{2m(r)}{r}} + r^2 d\Omega^2,
    \label{eq:gengrav}
\end{equation}
where $m(r)$ is the cumulative mass distribution and $\Phi(r)$ is the redshift function. 
The function $m(r)$ corresponds to the Arnowitt-Deser-Misner (ADM) mass $M$ only in the case of classical solutions \cite{Arnowitt:1962hi}. When short scale quantum effects are present, $m(r)$ assumes the more general definition of the integral of the mixed time-time component of the energy-momentum tensor $T_0^{\ 0}$. 
Due to symmetry constraints, the energy-momentum tensor can only have the form 
\begin{equation}
T_{\mu}^{\ \nu}= \text{diag}\left(-\rho(r),\ p_r(r),\ p_\perp(r),\ p_\perp(r)\right),
\label{eq:emt}
\end{equation} 
whose entries must decrease rapidly to interpolate a regular quantum region of constant energy density in the center with an outer classical vacuum. Quantum gravity comes into play by imposing a specific energy density profile. In the case of non-commutative geometry \cite{SmS04,SSN06,KoN10}, one finds
   \begin{equation}
\rho (r) = \frac{M}{(4\pi\theta)^{3/2}} e^{-\frac{r^2}{4\theta}},
\label{eq:rho}
\end{equation}
 which is constant at the origin, $\rho(0)\approx \rho_{\text{vac}}$, and vanishing at distances greater than the non-commutative scale $\sqrt{\theta}\sim l_{\rm P}$, as expected.
The solution of Einstein's equations requires an additional condition resulting from the choice of a particular equation of state. The most natural choice is $p_r(r)=-\rho(r)$, which implies the vanishing of the redshift function $\Phi(r)=0$, while the angular pressures are determined by the Tolman-Oppenheimer-Volkoff equation \cite{Tolman:1934za,Oppenheimer:1939ne},
\begin{equation}
    \frac{dp_r}{dr} = -\frac{1}{g_{00}}\frac{dg_{00}}{dr} (\rho + p_r) +\frac{2}{r} (p_\perp-p_r),
    \label{eq:tov}
\end{equation}
where
\begin{equation}
\frac{1}{g_{00}}\frac{dg_{00}}{dr}=\frac{m(r) + 4\pi r^3 p_r}{r^2 \left( 1-\frac{2m(r)}{r} \right)}.
\label{eq:g00}
\end{equation}

As mentioned in \ref{item:ii}), the above family of metrics \eqref{eq:gengrav} do not necessarily describe black holes.
In fact, horizons form only when the condition $r=2m(r)$ is satisfied, namely when the ADM mass $M$ is greater than or equal to a minimum mass $M_0$ with $M_0\simeq 1.9041\sqrt{\theta}$. This implies the existence of horizonless, self-gravitating objects similar to gravastars \cite{MaM23}, but with the following important differences.
The function $m(r)$ provides a smooth transition from the central quantum vacuum with  $T_{\mu\nu}\approx -\rho_{\text{vac}} \ g_{\mu\nu}$ to the outer classical vacuum with $T_{\mu\nu}= 0$.
In addition, the outer vacuum has a quantum perturbation due to the Gaussian tail of the energy density \eqref{eq:rho}, which makes these horizonless objects similar to a family of solutions called dark energy solitons or G-lumps \cite{Dym94,Dym25}.
According to Gliner, these objects result from the collapse of superdense matter into a superdense vacuum \cite{Gli66}. This vacuum state represents not only the final phase of the collapse, but also the initial state from which the universe undergoes expansion \cite{Gli70}.

\begin{figure}[h!]
\includegraphics[scale=.45]{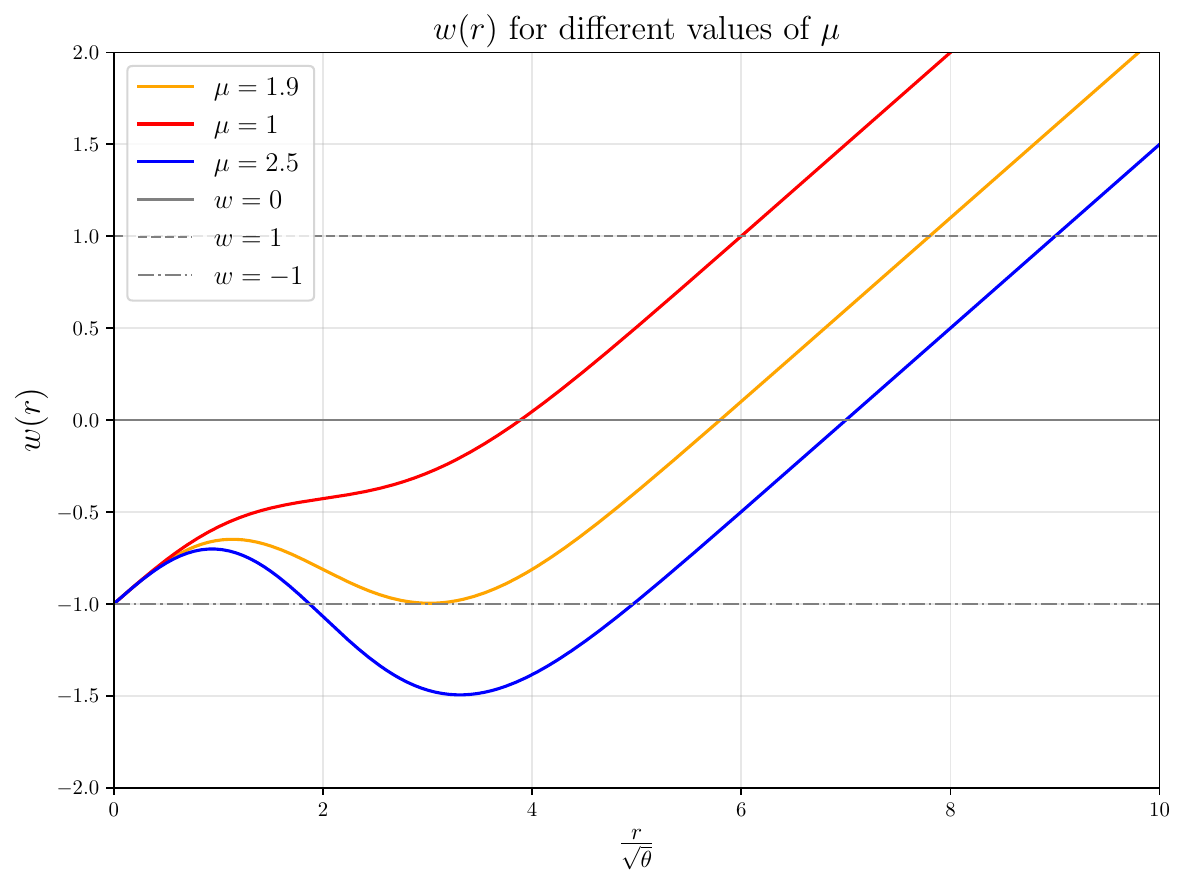}
\centering
\caption{Plots of \(w(r) = \frac{p_r(r)}{\rho(r)}\) versus \(\frac{r}{\sqrt{\theta}}\) for three different mass values \(\mu < \mu_0\). The curves are independent of \(\sqrt{\theta}\).}
\label{fig:w(r)}
\end{figure}

The metric \eqref{eq:gengrav} can, however, describe other objects with correspondingly different equations of state. 
For example, for $p_r(r)=-m(r)/(4\pi r^3)$, the metric describes a traversable wormhole \cite{Garattini:2008xz,Nicolini:2009gw}. For $p_r(r)=w(r)\rho(r)$ with
\begin{equation}
w(r) = \frac{p_r}{\rho} = - \left[1-\frac{r}{2\sqrt{\theta}} \left( 1- \frac{2m(r)}{r}\right)\right],
    \label{eq:eeooss}
\end{equation}
the metric describes a regular ``dirty black hole'' \cite{Nicolini:2009gw}, 
which is a black hole without curvature singularity that interacts with particles in its surroundings.
Due to the peculiar equation of state, the function $\Phi(r)$ is no longer vanishing and implies an increase of the red shift with respect to the corresponding regular black hole solutions with $\Phi=0$. 
Also for $\Phi\neq 0$ the metric \eqref{eq:gengrav} admits horizonless objects similar to the aforementioned G-lumps.
Nevertheless, for large $r$ the function $w(r)$ in \eqref{eq:eeooss} exceeds unity, which can only be accepted as a mathematical limit for pressure and density, both going to zero with the same power.

To avoid an unbounded growth of $w(r)$, in this work we assume a finite radius $R$ for the region where the energy-momentum tensor is nonvanishing.
The result is a new kind of spacetime with a local quantum vacuum, i.e., $w(0) \approx -1$ at the center, similar to a gravastar.
However, unlike a conventional gravastar, the function $w(r)$ grows as $r$ goes up to $R$, corresponding to matter shells with an increasingly rigid equation of state.
At the boundary, the radius $R$ is determined by solving the equation $w(R)=0$, the limit imposed by the requirement that the pressures vanish at the surface of the star.
A finite radius requires some adjustments in the energy density profile because the total mass must match the cumulative distribution at $R$ rather than at infinity. As a result in \eqref{eq:rho}, one has to do the substitution
\begin{equation}
M\longrightarrow \mu 
\end{equation}
where $\mu$ is a mass parameter chosen to satisfy the condition $M\equiv m(R)$, namely 
\begin{equation}
\mu = \frac{\Gamma\left(3/2\right) \ M}{\gamma\left(3/2; R^2/4\theta\right)} = \frac{\sqrt{\pi}}{2}\frac{M}{\gamma\left(3/2; R^2/4\theta\right)} >M
\end{equation}
Therefore the energy density is given by
\begin{equation}
  \rho(r)=  \begin{cases}
    \frac{\mu}{(4\pi\theta)^{3/2}} e^{-\frac{r^2}{4\theta}} \ \ \mathrm{for}\ r \leq R  \\
    0 \phantom{\frac{\mu}{(4\pi\theta)^{3/2}} e^{-\frac{r^2}{4\theta}}} \, \mathrm{for}\ r > R,
    \end{cases} \ ,
   \end{equation}
while the equation state is given by \eqref{eq:eeooss}, with the boundary condition $w(r)=0$ for $r\geq R$.
The spacetime is determined by calculating the cumulative mass distribution and the redshift function. From \eqref{eq:g00} and \eqref{eq:eeooss} one obtains
\begin{equation}
    \frac{d\Phi}{dr} = \begin{cases}
    \frac{4\pi r \left(\rho+p_r\right)}{1-2m(r)/r} = \frac{2\pi r^2}{\sqrt{\theta}} \rho(r) \ \ \mathrm{for}\ r \leq R  \\
    0 \phantom{\frac{4\pi r \left(\rho+p_r\right)}{1-2m(r)/r} = \frac{2\pi r^2}{\sqrt{\theta}} \rho(r)} \, \mathrm{for}\ r > R.
    \end{cases} \ ,
    \label{eq:red-shift}
\end{equation}
The solution to this equation is determined by requiring the boundary condition $\Phi(R) = 0$, or equivalently, $\Phi(r) \to 0$ as $r \to\infty$ and is given by
\begin{equation}
\Phi(r)= \begin{cases}-\frac{1}{2\sqrt{\theta}}\left[M-m(r)\right] \ \ \mathrm{for}\ r \leq R  \\
 0 \phantom{\frac{2\pi M}{\sqrt{\theta}}\left[M-m(r)\right]} \, \mathrm{for}\ r > R,
\end{cases} \ ,
\end{equation}

where
\begin{equation}
\label{eq:cummass}
m(r)=\frac{\gamma\left(3/2; r^2/4\theta\right)}{\gamma\left(3/2; R^2/4\theta\right)} M \quad \mathrm{for} \ r\leq R, 
\end{equation}
while it is equal to $M$ for $r>R$. 
 In summary, one ends up with an internal and an external solution, which must coincide at the boundary:

\begin{equation}
ds^2 = \begin{cases}
 -e^{-\frac{M}{\sqrt{\theta}} \left[1-m(r)/M\right] } \left( 1-\frac{2m(r)}{r}\right) dt^2 +\frac{dr^2}{1-\frac{2m(r)}{r}} + r^2 d\Omega^2 \\
 \quad \text{for } r\leq R \\ \\
    -\left( 1-\frac{2M}{r}\right) dt^2 +\frac{dr^2}{1-\frac{2M}{r}} + r^2 d\Omega^2 \\
    \quad \text{for } r> R.
\end{cases}
 \label{eq:w=0solution}
\end{equation}

Since the equation of state \eqref{eq:eeooss} guarantees that pressures do not diverge at any $r$, the singularity at the origin is removed and the curvature jump at $r=R$  is finite. 
To an outside observer, the spacetime appears to be that described by the Schwarzschild metric. However, there are important differences. 
In \eqref{eq:eeooss}, as $r$ increases from zero to $R$, the function $m(r)$ rises, causing the curve $w(r)$ to deviate from linear growth. The behavior of $w(r)$ depends on the value of the mass parameter $\mu$. For $\mu=\mu_0\simeq 1.9041\sqrt{\theta}$, the function $w(r)$ initially increases, reaches a maximum, and then decreases to a minimum value of $w=-1$ before growing again as $m(r)$ approaches its maximum value $M$ at the stellar surface $r = R$ (see Fig.~\ref{fig:w(r)}).
For higher values of $\mu$, the two horizons, $r_\pm$, that emerge from the equation $r = 2m(r)$ are loci where $w = -1$.
The equation $w(r)=0$ can be solved to find that 
\begin{equation}
R=2M+2\sqrt{\theta}.
\end{equation}
These results indicate that the horizons are located behind the stellar surface $r_-\leq r_+< R$. Furthermore, the radius of the star depends on two terms: a classical gravitational term, $2M$, and a term of quantum gravity origin, $2\sqrt{\theta}$. Unlike conventional stars, the limit of star radius for mass approaching zero is not zero but $R\approx 2\sqrt{\theta}$, due to the absence of infinite resolution over quantum spacetime. Another interesting characteristic of this star is that it is more compact than stars of uniform density. For $M > 8\sqrt{\theta}$, the conventional general relativity limit $R \geq (9/4) M$ is no longer satisfied \cite[p. 131]{Wal84}. This is due to the intrinsic quantum nature of the object.
However, for large total masses $M$, the minimum of the curve $w(r)$ deepens and enters the phantom matter regime, i.e., $w < -1$ \cite{Cal02}.  If we restrict the case so that pressures do not exceed their central values, then the minimum of $w(r)$ should not be that deep; that is, the mass $M$ cannot exceed $M_0\simeq 1.9025\sqrt{\theta}$ and the radius $R$ cannot exceed $R_0\equiv R(M_0)\simeq 5.8051\sqrt{\theta}$.

\subsection{Compact objects with quantum core}

There are other options besides the above model. Indeed, in the following, we distinguish three classes of configurations: (i) a pure quantum-vacuum star with pressure vanishing at a finite radius, and two quantum-core models (heavy and light), characterized by different dominance regimes of the envelope relative to the core.

One could think of the regular quantum vacuum at the center as representing the core of a larger star, with an envelope region providing the transition to the outer Schwarzschild geometry.

The density, $\rho$, is a continuous function that decreases from its central value until it reaches its minimum value at $R_1$:
\begin{equation}
  \rho(r)=  \begin{cases}
    \frac{\mu}{(4\pi\theta)^{3/2}} e^{-\frac{r^2}{4\theta}} \ \ \mathrm{for}\ r \leq R_1  \\
    \frac{\mu}{(4\pi\theta)^{3/2}} e^{-\frac{R_1^2}{4\theta}}\ \ \mathrm{for}\   R_1 <r \leq R_2\\
    0 \phantom{\frac{\mu}{(4\pi\theta)^{3/2}} e^{-\frac{r^2}{4\theta}}} \, \mathrm{for}\ r > R_2.
    \end{cases} \ ,
   \end{equation}
In the envelope, the density remains constant, maintaining the same value as at the exit from the core.  The simplest way to model the envelope is to assume conditions of a perfect fluid. To ensure a smooth transition from the core to the envelope, the radial and angular pressures must coincide at the core's boundary. Therefore, the condition $p_r(R_1)=p_\perp(R_1)$ is used to determine the radius $R_1$ of the core.

From \eqref{eq:tov}, the angular pressure reads:
\begin{equation}
p_\perp= w\rho +\frac{r}{2}\left(w^\prime \rho + w \rho^\prime \right)+\frac{\left( m+4\pi r^3 w\rho \right) \rho}{4\sqrt{\theta}}.
\end{equation}
Since the first term on the r.h.s. is the radial pressure, the other terms must vanish. We see that both $w^\prime=(1-2m^\prime)/2\sqrt{\theta}$ and $\rho^\prime=-r\rho/2\theta$ have negative contributions, while the rest of terms are positive. We can further simplify the equation by considering its leading terms. For $r\gtrsim R_0$, the density is  $\rho \sim 10^{-6}/\theta$, and $0\lesssim w\ll 1$, $m\lesssim 2\sqrt{\theta}$. Therefore, by keeping the leading terms, one finds that
\begin{equation}
p_\perp\approx w\rho +\frac{r\rho}{4\sqrt{\theta}}\left(1+\frac{3r}{\sqrt{\theta}}-\frac{r^2 }{2\theta} \right).
\end{equation}
Therefore the radius of the core is $R_1\approx 6.32\sqrt{\theta}$, which is determined by requiring that the term in parentheses vanishes in the previous equation.

The spacetime metric has the general form as follows:
\begin{equation}
ds^2 = 
\begin{cases}
 -e^{2\Phi_1(r)} \left( 1-\frac{2m_1(r)}{r}\right) dt^2 + \frac{dr^2}{1-\frac{2m_1(r)}{r}} + r^2 d\Omega^2 \\
 \quad \text{for } r \leq R_1 \\ \\
 -e^{2\Phi_2(r)} \left( 1-\frac{2m_2(r)}{r}\right) dt^2 + \frac{dr^2}{1-\frac{2m_2(r)}{r}} + r^2 d\Omega^2 \\
 \quad \text{for } R_1 < r \leq R_2 \\ \\
 -\left( 1-\frac{2M}{r}\right) dt^2 + \frac{dr^2}{1-\frac{2M}{r}} + r^2 d\Omega^2 \\
 \quad \text{for } r > R_2,
\end{cases}
\label{eq:evenlope-solution}
\end{equation}

where from \eqref{eq:cummass} $m_1(r)=M_1 \gamma(3/2; r^2/4\theta)/\gamma(3/2; R_1^2/4\theta)$, and
\begin{equation}
m_2(r)=M_1+\frac{4\pi \rho_1}{3}\left(r^3-R_1^3\right),
\end{equation}
with $\rho_1=\rho(R_1)$. Therefore, the total mass of the star is: 
\begin{equation}
M=M_1+\frac{4\pi \rho_1}{3}\left(R_2^3-R_1^3\right).
\end{equation}
To determine the redshift functions, the star's pressure must be derived from equation \eqref{eq:tov} in the envelope region. This can be done by recalling that $p_r(r)=p_\perp(r)\equiv p(r)$ for $R_1 < r \leq R_2$.
For later convenience, it is useful to introduce the length scale $L$, where $L^{-2} = 4\pi\rho_1/3$. As a result, the following equation applies to the envelope region: 
\begin{equation}
\begin{split}
\frac{dp}{dr} &= -\frac{\frac{r^3}{L^2}+M_1-\frac{R_1^3}{L^2}+4\pi r^3 p}{r\left(r- \frac{2r^3}{L^2}-2M_1+\frac{2R_1^3}{L^2}\right)}\left(\frac{3}{4\pi L^2}+p\right) \\
&\quad \text{for } R_1 < r \leq R_2,
\end{split}
\end{equation}

which can be approximated as
\begin{equation}
\begin{split}
\frac{dp}{dr} &\approx -\frac{r^3\left(\frac{1}{L^2}+4\pi p\right)+M_1}{r\left(r- \frac{2r^3}{L^2}-2M_1\right)}\left(\frac{3}{4\pi L^2}+p\right) \\
&\quad \text{for } R_1 < r \leq R_2.
\end{split}
\label{eq:tov2}
\end{equation}

The above approximation is valid because $R_1\ll L \sim 10^3\sqrt{\theta}$ and  $M_1 \simeq 2\sqrt{\theta}$. 
Therefore, the term $R_1^3/L^2$ can be neglected because it is much smaller than $M_1$.
To solve equation \eqref{eq:tov2}, two regimes can be considered:
\begin{enumerate}
    \item the large length scale regime where $r\gg(L^2M_1)^{1/3}$;
    \item the short length scale regime where ($r\ll(L^2M_1)^{1/3}$) and $(L^2M_1)^{1/3}\sim 2\times 10^2 \sqrt{\theta}$.
\end{enumerate}

\subsubsection{Large length scale regime}\label{sec:RGOs21}

\begin{figure}[h!]
\includegraphics[scale=.45]{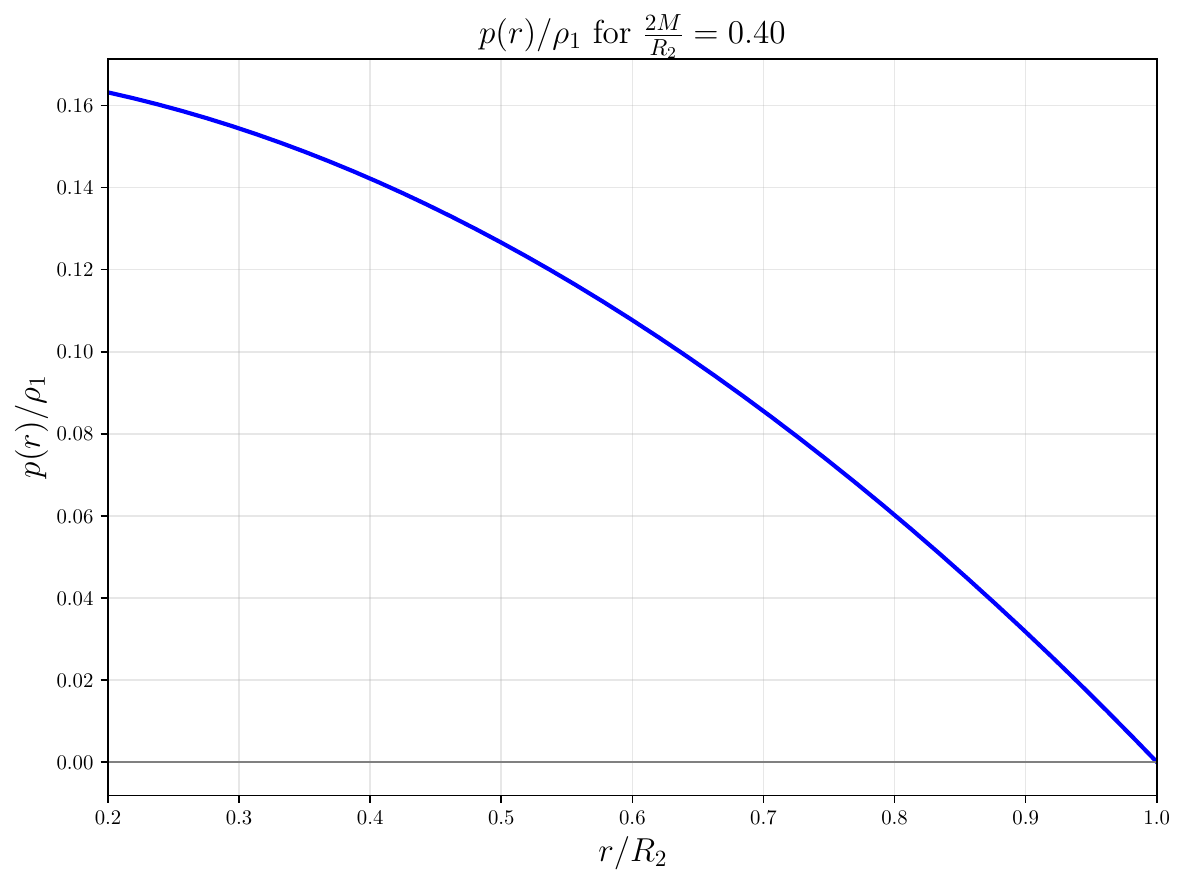}
\centering
\caption{Density profile of the envelope. Plot of the density $\rho(r)$ given by Eq.~\eqref{eq:penv}, in units of $\rho_1$, as a function of $r/R_1$. The parameters are chosen such that $R_1 \simeq 6.32\sqrt{\theta}$, $R_2 \simeq 30\sqrt{\theta}$, and $M \simeq 6\sqrt{\theta}$. This choice ensures $0.20 \leq r/R_2 \leq 1$ and $2M/R_2 \simeq 0.40$.
}
\label{fig:pr}
\end{figure}

The first case corresponds to a small core mass. Consequently, $M_1$ can be neglected in \eqref{eq:tov2}, and the star can be considered a classical object dominated by the envelope. In such a regime, one finds the conventional solution for pressure derived by Karl Schwarzschild
\cite{Sch16}
\begin{equation}
p(r)=\rho_1 \left( \frac{\sqrt{1-\frac{2M}{R_2}}-\sqrt{1-\frac{2Mr^2}{R_2^3}}}{\sqrt{1-\frac{2Mr^2}{R_2^3}}-3\sqrt{1-\frac{2M}{R_2}}}\right),
\label{eq:penv}
\end{equation}
where the total mass $M\equiv m_2(R_2)\simeq R_2^3/L^2\gg M_1$. Being \eqref{eq:penv} a monotonic decreasing function in $r$, the pressure in the envelope region  has its maximum value at $R_1$ and decreases as $r$ grows (see Fig. \ref{fig:pr}).

As requested for stellar stability, the pressure vanishes at $R_2$. Conversely, for $r\approx R_1\ll R_2$ the pressure becomes
\begin{equation}
p(R_1)\approx\rho_1 \left( \frac{1- \sqrt{1-\frac{2M}{R_2}}}{3\sqrt{1-\frac{2M}{R_2}}-1}\right).
\end{equation}
This leads to the conventional general relativity limit $M<(4/9)R_2$ for the start to exist. 
It is also important to ensure continuity of pressure between the core and envelope at $R_1$.  This implies that
\begin{equation}
w_1=\left( \frac{1- \sqrt{1-\frac{2M}{R_2}}}{3\sqrt{1-\frac{2M}{R_2}}-1}\right),
\label{eq:presscont}
\end{equation}
where, from \eqref{eq:eeooss},
\begin{equation}
w_1\equiv w(R_1)=-1+\frac{R_1}{2\sqrt{\theta}}- \frac{M_1}{\sqrt{\theta}}.
\end{equation}
For sufficiently low masses $M$, the r.h.s. in \eqref{eq:presscont} tends to $M/2R_2$. Therefore, the conditions for stellar mass and radius are as follows:
\begin{eqnarray}
M&=&\left(\frac{6}{\pi \rho_1}\right)^{1/2}\ w_1^{3/2}\\
R_2&=&\left(\frac{3}{2\pi \rho_1}\right)^{1/2}  \ w_1^{1/2}.
\end{eqnarray}
Note that both expressions depend on properties of the core only. The parameter $w_1$, which is approximately equal to 0.1, ensures that the system is physically reasonable by excluding large pressures at low densities.

\subsubsection{Short length scale regime}\label{sec:RGOs22}

The second case corresponds to a regime in which the core dominates the star's dynamics, and the envelope is too thin to have a significant effect. In this case, when $M\approx M_1$, \eqref{eq:tov2} becomes
\begin{equation}
\frac{dp}{dr}\approx -\frac{M}{r\left(r- 2M\right)}\left(\frac{3}{4\pi L^2}+p_r\right),  \label{eq:tov3}
\end{equation}
and the solution is:
\begin{equation}
p(r)=\rho_1\left(\sqrt{\frac{1-\frac{2M}{R_2}}{1-\frac{2M}{r}}}-1\right).
\end{equation}
The above expression ensures that the star's surface is stable, i.e., $p(R_2) = 0$. Conversely, the continuity condition at the interface between the core and the envelope implies
\begin{equation}
M=\left(\frac{R_2}{2}\right)\frac{1-(1+w_1)^2}{1-\left(\frac{R_2}{R_1}\right) (1+w_1)^2}.
\end{equation}

The mass exceeding the classical maximum of $(4/9)R_2$ is a signature of this compact object's quantum nature. Note that the radius $R_2$ must satisfy the condition $R_2 < R_1/w_1^2$, which is compatible with the regime under consideration.

\subsubsection{Red shift function}

The red shift function inside the core is 
\begin{equation}
\Phi_1(r)=\frac{m_1(r)}{2\sqrt{\theta}}+\Phi_1(0),
\end{equation}
where $\Phi_1(0)$ is an integration constant chosen so that  $\Phi_1(R_1) =  \Phi_2(R_1)$.
One again considers two regimes in the envelope: the light core, $m_2(r)\approx r^3/L^2 \gg M_1$, and the heavy core, $m_2 \approx M_1 \approx M$. In both cases, the pressure term can be neglected because $w(r) < w_1 \sim 0.1$ inside the envelope.
Therefore, the equations to solve are 
\begin{equation}
    \frac{d\Phi}{dr} \approx \begin{cases}
    \frac{4\pi r \rho}{1-2r^2/L^2}  \ \ \mathrm{for}\ r \gg (M_1L^2)^{1/3} \ (\textrm{light \ core})  \\
    \frac{4\pi r \rho}{1-2M/r}  \ \ \mathrm{for}\ r \ll (M_1L^2)^{1/3} \ (\textrm{heavy \ core}).
    \end{cases} \ ,
\end{equation}
The solutions are
\begin{widetext}
\begin{equation}
    \Phi(r) \approx \begin{cases}
    \Phi^{\mathrm{lc}}(R_1) +\frac{3}{2}\ln\left(\frac{R_2^3 -2 R_1^2 M}{R_2^3 -2 r^2 M}\right)  \ \ \mathrm{for}\ r \gg (M_1L^2)^{1/3} \ (\textrm{light \ core})  \\
    \Phi^{\mathrm{hc}}(R_1) +2\pi\rho_1\left[r^2+4Mr-4MR_1-R_1^2+4M\ln\left(\frac{r-2M}{R_1-2M}\right)\right]   \ \ \mathrm{for}\ r \ll (M_1L^2)^{1/3} \ (\textrm{heavy \ core})
    \end{cases} \ ,
\end{equation}
    
\end{widetext}

where the integration constants $\Phi^{\mathrm{lc}}(R_1)$ and $\Phi^{\mathrm{hc}}(R_1)$ can be chosen to satisfy the condition, $\Phi_2(R_2)=0$. 

\section{Discussion of predictions}\label{sec:models}

In the previous section, we developed three distinct theoretical models for compact stellar objects within the framework of noncommutative geometry:

\begin{enumerate}
\item \textbf{Quantum vacuum star}: A pure quantum vacuum configuration with a uniform equation of state throughout the entire stellar interior (Section \ref{sec:RGOs}).
\item \textbf{Heavy quantum core star}: A configuration featuring a dense quantum core surrounded by a light envelope, characterized by a large mass-to-radius ratio in the envelope region (Section \ref{sec:RGOs21}, large length scale regime).
\item \textbf{Light quantum core star}: A configuration with a quantum core surrounded by a more massive envelope, characterized by a smaller mass-to-radius ratio (Section \ref{sec:RGOs22}, short length scale regime).
\end{enumerate}

Each of these three models is characterized by a single fundamental parameter, $\sqrt{\theta}$, which represents the length scale at which noncommutative geometry effects become significant. Once $\sqrt{\theta}$ is specified, all other physical parameters---including the stellar radius, mass, and central density---are uniquely determined by the model equations. These three stellar configurations respond differently to the choice of the noncommutative scale, because their compactness and density gradients scale with $\sqrt{\theta}$ in distinct ways. Therefore, exploring different regimes of the parameter allows us to map how quantum effects propagate from the core to the outer geometry

To explore the phenomenological implications of these models, we now examine their predictions across different energy regimes by varying the parameter $\Lambda \equiv 1/\sqrt{\theta}$. This approach allows us to investigate four distinct physical scenarios:

\begin{enumerate}
\item \textbf{Planckian scenario} ($\Lambda = m_\mathrm{P}$, $\sqrt{\theta} \simeq 1.62 \times 10^{-35}$ m): The regime where quantum gravity effects are expected to be maximal.
\item \textbf{TeV scenario} ($\Lambda \sim 10^{-15}$, $\sqrt{\theta} \sim 10^{-20}$ m): The scale corresponding to the current experimental limits from particle physics.
\item \textbf{Submillimeter scenario} ($\Lambda \sim 10^{-31}$, $\sqrt{\theta} \sim 10^{-4}$m): The scale probed by Cavendish-like gravity experiments, representing the smallest length at which gravity has been directly tested.
\item \textbf{Nuclear scenario} ($\Lambda \sim 10^{-38}$, $\sqrt{\theta} \sim 10^{2}$ m): A phenomenological regime where the models can describe objects with astrophysically relevant masses and radii comparable to neutron stars.
\end{enumerate}

Tables \ref{tab:quantum-vacuum-star}, \ref{tab:heavy-core-star}, and \ref{tab:light-core-star} present the detailed predictions for each of the three models across these four scenarios, showing how the fundamental stellar parameters scale with the noncommutative parameter $\sqrt{\theta}$. In what follows, we discuss each scenario in detail, exploring the physical characteristics and observational prospects of the resulting stellar configurations.

The primary challenge lies in selecting a realistic regime for $\sqrt{\theta}$. A conventional approach is to adopt the Planck scale, where $\sqrt{\theta} \simeq 1.62 \times 10^{-35}$ m, leading to the Planckian scenario. However, this choice leads to stellar models representing extremely minute objects that would be astronomically challenging to detect. At this scale, the objects' dimensions would approximate both their gravitational radius and Compton wavelength, suggesting a regime dominated by concurrent strong gravitational field and quantum mechanical effects.
The resulting stellar configurations would exhibit a density approximating unity in Planck units---a physical state that lies beyond our current understanding of matter under extreme pressure conditions. Conceptually, these Planckian objects would resemble compact, stringy matter packets on a microscopic scale, bearing similarities to theoretical constructs within the fuzzball paradigm \cite{Mat05}.

Reducing the energy scale reveals new stellar scenarios potentially observable by modern telescopes. For an energy scale $\Lambda\equiv 1/\sqrt{\theta} < 1$, the star's radius and mass scale as $1/\Lambda$, while density decreases quadratically, $\sim \Lambda^{2}$. Notably, gravitational radius $r_\mathrm{g}$ and Compton wavelength $\lambda$ exhibit opposite $\Lambda$ dependencies, with classical gravity becoming the dominant factor of the star dynamics below the Planck scale. Please note that, for simplicity, we set $\rho_0 = 1$ at the Planck scale by neglecting the extra factor $\frac{1}{(4\pi)^{3/2}} \simeq 2.24 \times 10^{-2}$, which would slightly enhance the masses and radii displayed.  

In principle, the energy scale $\Lambda$ can be lowered indefinitely; however, experimental limits from particle physics suggest that a conservative lower bound for the scale at which noncommutative-geometry effects might emerge is around 10 TeV (TeV scenario). Despite this, each of the three models remains constrained to sizes in the microscopic realm, meaning they are smaller than known elementary particles. As a result, in the absence of further hypotheses, such objects can only be relevant for terascale quantum gravity, but in contrast to elementary particles, their nature is purely classical, as their Compton wavelength is smaller than their gravitational radius, which in turn is smaller than their actual radius. Consequently, they would correspond to extreme gravitational field objects capable of disturbing quantum fields and providing intense vacuum polarization effects similar to black holes within the framework of the semiclassical gravity regime \cite{BiD84}. 

To further decrease $\Lambda$, one could circumvent the constraints from particle physics by assuming that, while the Standard Model fields remain unaffected, noncommutativity or other non-classical effects can modify gravity at larger scales. In this scenario, for the sake of consistency, one can consider the smallest length scale at which gravity has been experimentally probed in Cavendish-like experiments as the upper limit at which the sought effects may manifest \cite{Hoyle:2004cw,Adelberger:2006dh,Adelberger:2009zz,Long:2003dx}. This scale corresponds to $44 \, \mu\text{m}$, resulting in a value of $\Lambda \sim 10^{-31}$. In this regime, the masses fall within the range between the mass of the Moon and the mass of the Earth, depending on the model. However, their size is on the order of millimeters, slightly larger than their gravitational radius due to their extreme compactness, making them exceedingly difficult to detect astrophysically.

A more radical and phenomenologically promising approach would be to set the mass of the star in the regime of solar mass, or equivalently, to define the core density governed by energies between the MeV and the GeV. In this context, quantum gravity can no longer serve as the primary motivation for star model development, as the noncommutative parameter $\sqrt{\theta}$ becomes on the order of $10^2 \, \text{m}$. Nevertheless, the overall result remains intriguing because the three metrics we derived still represent star models with a quantum vacuum, showcasing extreme compactness while maintaining reasonable mass and radius parameters suitable for describing neutron stars, or, more speculatively, quark matter stars \cite{Lattimer:2004sa}—refer to the column titled ``Nuclear'' in Tab. \ref{tab:quantum-vacuum-star}-\ref{tab:heavy-core-star}-\ref{tab:light-core-star}.

\begin{widetext}
\begin{table}[ht!]
\centering
\hspace*{1.7cm}
\begingroup
\setlength{\tabcolsep}{4pt} 
\renewcommand{\arraystretch}{0.95} 
\small
\begin{tabular}{|c|c|c|c|c|c|}
\hline
& $\Lambda$ & Planckian ($\Lambda = m_\mathrm{P}$) & 10 TeV ($\Lambda\sim 10^{-15}$) & 44\,\textmu m ($\Lambda\sim 10^{-31}$) & Nuclear ($\Lambda\sim 10^{-38}$) \\
\hline
$\sqrt{\theta}$ & $\phantom{.}1/\Lambda$ & 1 & $1 \times 10^{-20}$ m & $1 \times 10^{-4}$ m  & $5 \times 10^{2}$ m  \\
\hline
$\rho_0$ & $\phantom{.}\Lambda^{2}$ & 1 & $1\times 10^{-30}$\phantom{ m}   & $1\times 10^{-62}\sim \textrm{TeV}^4$ &  $4\times 10^{-76}\sim \textrm{GeV}^4$ \\
\hline
$M$ & $\phantom{.}2/\Lambda$ & 2 & $4\times 10^{7}$ kg\phantom{m}   & $4\times 10^{23}$ kg &  $2\times 10^{30}$ kg $\simeq M_\odot$  \\
\hline
$R$ & $\phantom{.}6/\Lambda$ & 6 & $1\times 10^{-19}$ m  & $1\times 10^{-3}$ m  & $5\times 10^{3}$ m  \\
\hline
$r_\mathrm{g}$ & $\phantom{.}4/\Lambda$ & 4 & $6\times 10^{-20}$ m  & $6\times 10^{-4}$ m  & $3\times 10^{3}$ m \\
\hline
$\lambda$ & $.5\Lambda$ & .5 & $8\times 10^{-51}$ m  & \phantom{.}$8\times 10^{-67}$ m  &  \phantom{.ii}$3\times 10^{-80}$ m  \\
\hline
\end{tabular}
\endgroup
\caption{Quantum vacuum star.}
\label{tab:quantum-vacuum-star}
\end{table}

\begin{table}[ht!]
\centering
\begingroup
\setlength{\tabcolsep}{4pt} 
\renewcommand{\arraystretch}{0.95} 
\small
\begin{tabular}{|c|c|c|c|c|c|}
\hline
& $\Lambda$ & Planckian ($\Lambda = m_\mathrm{P}$) & 10 TeV ($\Lambda= \sim 10^{-15}$) & 44\,\textmu m ($\Lambda \sim10^{-31}$) & Nuclear $(\Lambda\sim 10^{-38})$ \\
\hline
$\sqrt{\theta}$ & $\phantom{.}1/\Lambda$ & 1 & $1 \times 10^{-20}$ m & $1 \times 10^{-4}$ m  & $5 \times 10^{2}$ m  \\
\hline
$\rho_0$ & $\phantom{.}\Lambda^{2}$ & 1 & $1\times 10^{-30}$\phantom{ m}   & $1\times 10^{-62} \sim \textrm{TeV}^4$ &  $4\times 10^{-76}\sim \textrm{GeV}^4$ \\
\hline
$\rho_1$ & $5\times 10^{-5}\Lambda^{2}$ & $5\times 10^{-5}$ & $5\times 10^{-35}$\phantom{ m}   & $5\times 10^{-67}\sim (84 \ \mathrm{GeV})^4$  & $5\times 10^{-81}\sim (84 \ \mathrm{MeV})^4$ \\
\hline
$M$ & $\phantom{.}2/\Lambda$ & 2 & $4\times 10^7$kg\phantom{ m}   & $4\times 10^{23}$kg\phantom{ m} &  $2\times 10^{30}$ kg $\simeq M_\odot$ \\
\hline
$R_2$ & $\phantom{.}7/\Lambda$ & 7 & $1\times 10^{-19}$ m  & $1\times 10^{-3}$ m & $5\times 10^{3}$ m  \\
\hline
$r_\mathrm{g}$ & $\phantom{.}4/\Lambda$ & 4 & $6\times 10^{-20}$ m  & $6\times 10^{-4}$ m  & $3\times 10^{3}$ m  \\ 
\hline
$\lambda$ & $.5\Lambda$ & .5 &$8\times 10^{-51}$ m  & \phantom{.}$8\times 10^{-67}$ m  &  \phantom{.ii}$3\times 10^{-80}$ m \\
\hline
\end{tabular}
\endgroup
\caption{Heavy quantum core star.}
\label{tab:heavy-core-star}
\end{table}

\begin{table}[ht!]
\centering
\begingroup
\setlength{\tabcolsep}{4pt} 
\renewcommand{\arraystretch}{0.95} 
\small
\begin{tabular}{|c|c|c|c|c|c|}
\hline
& $\Lambda$ & Planckian ($\Lambda = m_\mathrm{P}$) & 10 TeV ($\Lambda \sim 10^{-15}$) & 44\,\textmu m ($\Lambda \sim 10^{-31}$) & Nuclear $(\Lambda\sim 10^{-38})$ \\
\hline
$\sqrt{\theta}$ & $\phantom{.}1/\Lambda$ & 1 & $1 \times 10^{-20}$ m & $1 \times 10^{-4}$ m  & $5 \times 10^{2}$ m  \\
\hline
$\rho_0$ & $\phantom{.}\Lambda^{2}$ & 1 & $1\times 10^{-30}$\phantom{ m}   & $1\times 10^{-62}$\phantom{m}   &  $4\times 10^{-76}$\phantom{.} \\
\hline
$\rho_1$ & $5\times 10^{-5}\Lambda^{2}$ & $5\times 10^{-5}$ & $5\times 10^{-35}$\phantom{ m}   & $5\times 10^{-67}\sim (84 \ \mathrm{GeV})^4$  & $5\times 10^{-81}\sim (84 \ \mathrm{MeV})^4$ \\
\hline
$M$ & $\phantom{.}6/\Lambda$ & 6 &   $1\times 10^8$ kg\phantom{ m}  & $1\times 10^{24}$ kg &   $6\times 10^{30}$ kg $\simeq 3 M_\odot$  \\ 
\hline
$R_2$ & $\phantom{.}3/\Lambda\times 10^1$ & $3\times 10^1$ & $8\times 10^{-19}$ m  & $8\times 10^{-3}$ m &   $4\times 10^{4}$ m \\
\hline
$r_\mathrm{g}$ & $\phantom{.}4/\Lambda$ & 4 &  $2 \times 10^{-19} \, \text{m}$ & $2 \times 10^{-3} \, \text{m}$ & $9 \times 10^{3} \, \text{m}$ \\
\hline
$\lambda$ & $.5\Lambda$ & .5 & $3 \times 10^{-51} \, \text{m}$ & $3 \times 10^{-67} \, \text{m}$ & $1 \times 10^{-80} \, \text{m}$ \\ 
\hline
\end{tabular}
\endgroup
\caption{Light quantum core star.}
\label{tab:light-core-star}
\end{table}
\end{widetext}

\section{Conclusions}
\label{sec:conclusions}

In this work, we started from the horizonless solutions of noncommutative geometry to construct analytical solutions for compact stellar objects. We derived three distinct classes of metrics: a pure quantum vacuum star with a uniform equation of state throughout, and two models featuring a quantum core surrounded by an external envelope—distinguished by their different mass-to-radius ratios (heavy and light quantum core configurations). These analytical solutions extend the previous NCG framework beyond black holes to encompass a richer phenomenology of compact objects and how quantum gravity can manifest within them.
 These models demonstrate extreme compactness and feature a quantum core, effectively avoiding curvature singularities. This approach provides valuable insights into the fundamental nature of gravitational collapse at quantum scales.

The primary parameter governing these models is the noncommutative parameter, which can span from the TeV scale to the Planck scale. This inclusion provides greater flexibility in exploring various astrophysical scenarios, paralleling the role of $\alpha'$ in string theory. We have also examined lower energy regimes for this parameter, including the scale of submillimeter gravity experiments and the MeV scale. This analysis highlights the diverse potential of these models to bridge the gap between high-energy physics and observable phenomena.

Despite the intriguing characteristics of these stellar models, genuine, pure Planckian effects pose substantial challenges for their detection, confirming the limitations highlighted in our introduction regarding stellar astrophysics. However, we also identified that lower energy scales could reveal exotic objects with masses comparable to those of Earth, which are absent in classical general relativity. The detection of such objects could signify the first experimental evidence of non-classical gravity effects, thus addressing the fundamental question of how quantum gravity could be understood through astrophysical observations.
While these exotic objects are predicted to have small radii and significant mass, their actual detection remains an open question for future work. This emphasizes the need for innovative observational strategies to identify these compact objects. A further natural direction of this work concerns the observational prospects of the quantum-supported compact 
objects introduced here. Although these configurations share several features with ultra-compact stars and 
black-hole mimickers, their internal structure-dominated by a finite-size quantum core-and their lack of an 
event horizon imply measurable deviations that future multi-messenger facilities may be able to probe.
From the electromagnetic side, stellar-mass configurations can be tested against high-precision X-ray observations 
performed by NICER \cite{NICER,NICERa} and, in the near future, by ATHENA \cite{ATHENA} and eXTP \cite{eXTP}. The modified redshift function and the distinct 
density profile of our models can affect pulse-profile modelling, thermal emission, and cooling curves, introducing 
departures from standard neutron-star equations of state at the few-percent level. These effects could provide 
upper bounds on $\sqrt{\theta}$ in regimes ranging from nuclear densities to submillimeter gravity scales. At higher compactness, the absence of an event horizon influences the geometry of null geodesics and the structure 
of photon orbits. This makes these solutions particularly relevant for horizon-scale imaging with the Event Horizon 
Telescope and its upgrades \cite{2019EHTpaper1,2022EHTpaper1,2022EHTpaper6}. Modifications in the size of the shadow, the behaviour of 
higher-order photon rings, and the polarimetric structure of the emission may provide discriminants from the 
Schwarzschild geometry. The quasi-normal mode spectrum of the spacetime is also expected to differ, opening the possibility of identifying signatures of horizonless ultra-compact objects through ringdown modelling and time-domain reconstruction of photon-orbit structures \cite{Cardoso:2019rvt}. Gravitational-wave observations offer an equally promising avenue. For stellar-mass systems, LIGO–Virgo–KAGRA can 
constrain tidal Love numbers, which are non-zero for horizonless compact objects, in contrast to classical black holes \cite{2017GWS,Hinderer2007mb}. In the intermediate-mass regime, LISA will be able to detect extreme mass ratio inspirals (EMRIs) whose orbital evolution probes the full multipolar structure of the central object, including potential resonances 
generated by a quantum core \cite{Babak2017tow,Barack2003fp}. Next-generation observatories such as the Einstein Telescope  \cite{ET2019dnz} and Cosmic Explorer \cite{Reitze2019iox} may also detect post-merger signals and possible “echoes’’ associated with the absence of an event horizon-features that would be strong indicators of the type of configurations studied here \cite{Cardoso2016rao}.
Another scenario arises at lower values of $\Lambda$, where the models predict Earth-to Moon-mass compact objects with radii only slightly above their gravitational radius. These could leave detectable imprints in gravitational 
microlensing surveys such as Euclid, the Roman Space Telescope, and LSST. Their nearly point-like lensing behaviour, 
combined with finite-size corrections and potential surface emission under accretion, may help distinguish them 
from classical primordial black holes or other compact dark matter candidates. On the theoretical side, several developments are essential for establishing the robustness of these solutions. 
A key issue is the dynamical formation mechanism: whether such objects can emerge from realistic gravitational 
collapse, how the presence of a quantum core alters the collapse endpoint, and under which conditions they compete 
with classical black-hole formation. Stability under radial and non-radial perturbations must also be addressed, 
as the presence of a finite-radius quantum region may lead to characteristic oscillation modes or internal 
instabilities. Furthermore, extending the present analysis to rotating configurations is expected to reveal 
non-Kerr multipolar structures and modified frame-dragging profiles, which could act as observational discriminants 
in both X-ray and gravitational-wave observations.
Overall, these considerations indicate that the metrics derived in this work form a fertile basis for a 
phenomenology of quantum-corrected compact objects. By treating the noncommutative parameter as an effective scale 
governing the onset of a local de Sitter core, these models can be systematically confronted with present and 
upcoming data across the electromagnetic and gravitational-wave spectra. In this sense, they offer a promising 
opportunity to explore how quantum gravity might leave astrophysical signatures accessible to current and future 
experiments.

Finally, the most promising approach involves using the metrics developed in this study to conduct comparative tests against known neutron stars. We propose that the noncommutative parameters can be utilized as a general scale governing a macroscopic de Sitter region around the center, offering a unique opportunity to explore the interplay between quantum effects in gravity and astrophysical phenomena in future investigations. This approach could lead to a deeper understanding of the universe's earliest conditions and the fundamental laws of physics.

\begin{acknowledgments}
The research presented in this paper is the outcome of a summer study for graduate students in 2021, part of an Erasmus exchange between the University of Naples and Goethe University Frankfurt \cite{Sir22}.

S.S.S. wishes to thank the “Iniziativa Specifica InDark” of the INFN (Italian National Institute for Nuclear Physics) and Goethe University Frankfurt. Additionally, S.S.S. acknowledges the support from the PhD Program in Space Science and Technology at the University of Trento, funded by the European Union under the “NextGeneration EU” initiative—Mission 4 “Education and Research,” Component 2 “From Research to Business,” Investment 3.3.

P.N. wishes to thank the GNFM (Italian National Group for Mathematical Physics) and the “Iniziativa Specifica FLAG” of the INFN.

M.D.L. is grateful for the support from the “Iniziativa Specifica TEONGRAV and ET” of the INFN.
\end{acknowledgments}

\bibliographystyle{apsrev4-1}  
\bibliography{reference}

\end{document}